\documentclass[%
 reprint,
]{revtex4-1}

\usepackage{graphicx}
\usepackage{dcolumn}
\usepackage{bm}

\begin{document}

\preprint{APS/123-QED}

\title{Link Prediction with Node Clustering Coefficient}

\author{Zhihao Wu }
\email{zhwu@bjtu.edu.cn}
 
\author{Youfang Lin}
\author{Jing Wang}
\affiliation{%
Beijing Key Lab of Traffic Data Analysis and Mining,\\
School of Computer and Information Technology,\\
Beijing Jiaotong Univerisy, Beijing 100044, People's Republic of China
}%
\author{Steve Gregory}
\affiliation{%
 Department of Computer Science,\\
University of Bristol, Bristol, BS8 1UB, United Kingdom
}%


\begin{abstract}
Predicting missing links in incomplete complex networks efficiently and accurately is still a challenging problem. The recently proposed Cannistrai-Alanis-Ravai (CAR) index shows the power of local link/triangle information in improving link-prediction accuracy. Inspired by the idea of employing local link/triangle information, we propose a new similarity index with more local structure information. In our method, local link/triangle structure information can be conveyed by clustering coefficient of common-neighbors directly. The reason why clustering coefficient has good effectiveness in estimating the contribution of a common-neighbor is that it employs links existing between neighbors of a common-neighbor and these links have the same structural position with the candidate link to this common-neighbor. In our experiments, three estimators: precision, AUP and AUC are used to evaluate the accuracy of link prediction algorithms. Experimental results on ten tested networks drawn from various fields show that our new index is more effective in predicting missing links than CAR index, especially for networks with low correlation between number of common-neighbors and number of links between common-neighbors.\\
\\
\end{abstract}

\maketitle


\section{INTRODUCTION}

Complex network has shown its significant power in modeling and analyzing a wide range of complex systems, such as social, biological and information systems, and the study of complex networks has attracted increasing attention and becomes a popular tool in many different branches of science \cite{albert2002statistical,boccaletti2006complex,costa2007characterization,bianconi2009assessing,shen2014collective}. Prediction is one of the key problems in various research and application fields. Link prediction in complex networks aims at estimating the likelihood of the existence of a link between two nodes, and it has many applications in different fields. For example, predicting whether two users know each other can be used to recommend new friends in Social Networking Sites, and in the field of biology, accurate prediction of protein-protein interaction has great value to sharply reduce the experimental costs. Some researchers also applied the link prediction algorithms in partially labeled networks for prediction of protein functions or research areas of scientific publication \cite{holme2005role,gallagher2008using}. In addition, the study of link prediction is closely related to the problem of network evolving mechanisms \cite{wang2012evaluating,zhang2015Measuring}. Qianming Zhang and Tao Zhou et. al. employed link prediction methods to evaluate network models and attained better results than some classical models \cite{wang2012evaluating}. Recently, through measuring multiple evolution mechanisms of complex networks, they found the evolution of most networks is affected by both popularity and clustering at the same time, but with quite different weights \cite{zhang2015Measuring}.

Many link prediction methods have been proposed under different backgrounds in recent years \cite{liben2007link,lu2011link}. In this paper, we only focus on similarity-based methods using topology structural information. The basic assumption for this kind of link prediction methods is that two nodes are more likely to have a link if they are similar to each other. Therefore, the key problem is to define proper similarity measures between nodes. Some methods combine many factors to define the similarity between nodes, such as attributes of nodes and links and structural information. One group of similarity indices is based solely on the network structure. The simplest one is PA index \cite{newman2001clustering}, which is defined as the product of degrees of two seed nodes. Common-Neighbor (CN) \cite{lorrain1971structural} counts the number of common-neighbors and Jaccard index (JC) \cite{jaccard1901distribution} is a normalization of CN. To get better resolution, Adamic-Adar (AA) \cite{adamic2003friends} and Resource Allocation (RA)\cite{zhou2009predicting} are defined by employing the degree information of common-neighbors. Recently, a new index, called Cannistrai-Alanis-Ravai (CAR) \cite{cannistraci2013link}, is proposed by Cannistraci et al. Their main point is that link information of common-neighbors is useful but still noisy. They find level-2 links, i.e. links between common-neighbors, are more valuable and can be used to improve most classical Node-Neighborhood-based similarity indices. The above methods are all local measures, and to pursue higher prediction precision some global and quasi-local methods are also proposed, such as Katz \cite{katz1953new}, SimRank \cite{jeh2002simrank}, Hitting Time \cite{gobel1974random}, Average Commute Time \cite{fouss2007random}, Local Path \cite{lu2009similarity}, Transferring Similarity \cite{sun2009information}, Matrix Forest Index \cite{chebotarev1997matrix} and so on. Obviously, considering more information and features in prediction methods may cause more time and space costs.

Besides, there are also some more complex models and methods to solve the link prediction problem. Clauset et al. proposed an algorithm based on the hierarchical network structure, which gives good predictions for the networks with hierarchical structures \cite{clauset2008hierarchical,redner2008networks}. Guimera et al. solved this problem using stochastic block model \cite{guimera2009missing}. Recently, Linyuan L{\"u} et al. proposed a concept of structural consistency, which could reflect the inherent link predictability of a network, and they also proposed a structural perturbation method for link prediction, which is more accurate and robust than the state-of-the art method \cite{lu2015toward}. Although the above methods can attain better results than most Node-Neighborhood-based methods, they are hard to be applied to large networks.

Heretofore, efficient link prediction is still a big challenge. In our opinion, local methods are still good candidates for solving link prediction problem in large networks. Some results have shown that community/cluster structures can help improve the performance of link prediction \cite{liu2013correlations,cannistraci2013link}. Some researchers directly combine the communities detected by various community detection algorithms with some similarity indices, and show that cluster information can improve link prediction algorithms a lot in some cases \cite{yan2012finding,soundarajan2012using}. This kind of methods relies on the community detection algorithms, but there are lots of different algorithms. How to choose a proper algorithm is still not very clear.

In this paper, we present a new similarity index, called CCLP (Clustering Coefficient for Link Prediction), which employs more local link/triangle structure information than CAR index, but costs less computational time. The key idea of our method is to exploit the value of links between other neighbors of common-neighbors, except seed nodes and common-neighbor nodes, and they can be efficiently conveyed by using clustering coefficient of common-neighbors. Some related literatures also suggest that clustering coefficient has some relations with the problem of link prediciton problem \cite{feng2012link,liu2011link}. The experimental results on 10 networks from five various fields show that our new method performs better than CAR index on networks with not very high $LCP_{-corr}$ and is more efficient.

\section{METHODS}

\subsection{Definition}
Considering an unweighted undirected simple network $G(V,E)$, where $V$ is the set of nodes and $E$ is the set of links. For each pair of nodes, $x,y\in V$, we assign a score to the pair of seed nodes. All the nonexistent links are sorted in decreasing order according to their scores, and the links in the top are most likely to exist. The common-used framework always sets the similarity to the score, so the higher score means the higher similarity, and vice versa. The definitions of similarity indices mentioned in this paper are as follows.

\textbf{Preferential Attachment (PA).} 

\begin{equation}
PA(x,y) = |\Gamma(x)| \cdot |\Gamma(y)|
\label{eq1}
\end{equation}

where $\Gamma(x)$ denotes the set of neighbors of node $x$ and $|A|$ is the number of elements in set A.

\textbf{Common Neighbor (CN).} 

\begin{equation}
CN(x,y) = |\Gamma(x) \cap \Gamma(y)|
\label{eq2}
\end{equation}

\textbf{Jaccard (JC).}

\begin{equation}
JC(x,y) = \frac{|\Gamma(x) \cap \Gamma(y)|}{|\Gamma(x) \cup \Gamma(y)|}
\label{eq3}
\end{equation}

\textbf{Adamic-Adar (AA).}

\begin{equation}
AA(x,y) = \sum_{z\in\Gamma(x)\cap\Gamma(y)}\frac{1}{log(k_{z})}
\label{eq4}
\end{equation}
where $k_z$ is the degree of node $z$.

\textbf{Resource Allocation (RA).}

\begin{equation}
RA(x,y) = \sum_{z\in\Gamma(x)\cap\Gamma(y)}\frac{1}{k_{z}}
\label{eq5}
\end{equation}

\textbf{CAR.} 

\begin{equation}
CAR(x,y) = CN(x,y) \cdot LCL(x,y)
\label{eq6}
\end{equation}

where LCL(x,y) is the number of links between common-neighbors.

\subsection{Similarity index based on clustering coefficient}

\begin{figure*}[ht]
\centering
\includegraphics[width=\linewidth]{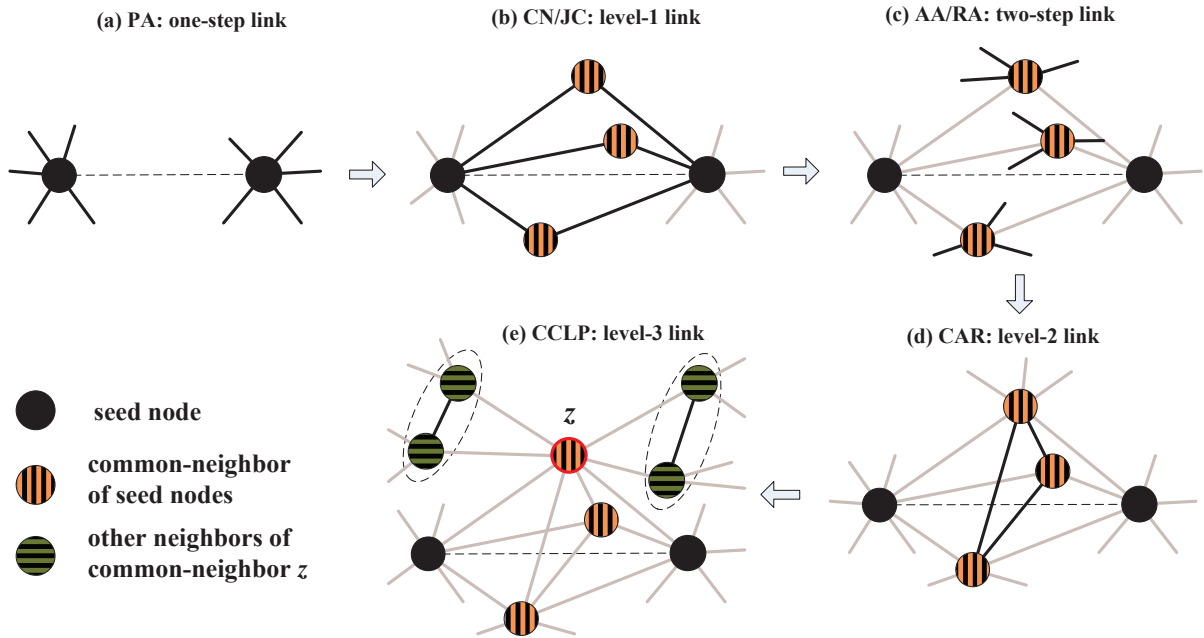}
\caption{\textbf{The illustration of local similarity indices.} (a) PA index is the simplest similarity index. Only one-step link information, which indicates all links connect to seed nodes, is used. (b) In one-step link information, level-1 links, which are links between seed node and common-neighbor, are found more valuable, and are employed by CN and JC indices. (c) Further, AA and RA indices prove two-step link information conveyed by common-neighbor can be used to increase the discriminative resolution of similarity index. (d) Cannistraci et al. pointed out that two-step link information is valuable but still noisy, and they proposed to employ level-2 links, which are links between common-neighbors, to improve classical similarity indices. (e) In this paper, we want to show the value of level-3 links, which are, for example, links between other neighbors of common-neighbor $z$. Other neighbors here mean that they are neither seed nodes nor common-neighbor nodes. CCLP index conveys level-3 links by using clustering coefficient of common-neighbors. For example, to calculate the clustering coefficient of node $z$, we need to count the number of links between neighbors of node $z$, because this number is just the number of triangles passing through node $z$. The most important reason why we consider level-3 links is because they have similar structural position with the candidate link corresponding to each common-neighbor. }
\label{fig:progress}
\end{figure*} 

Certainly, complicated techniques are able to give some insights into the mechanism of link formation, network evolution, and even the link predictability . However, some researchers also pointed out that these elegant techniques are at the moment mere proof of concepts rather than concrete methods to apply on real problems \cite{cannistraci2013link}. Except the problem of tuning parameters, they are not efficient enough either, because they can only predict missing links on very small networks (with few hundreds of nodes). Nevertheless, at the current age of big data, we always need to process very large networks. 

For these reasons, efficient and parameter-free methods still deserve to be further studied. Node-Neighborhood-based approaches are commonly used in both research and application. In these methods, classical ones, such as PA, CN, JC, AA and RA, only focus on node degree and common-neighbors. Recently Cannistraci et al. proposed a new index, called CAR, with the concept of local-community-paradigm. They also extended other Node-Neighborhood-based methods with LCL, and showed us that most classical Node-Neighborhood-based methods can be improved a lot combining LCL. 

LCL counts the number of links between common-neighbors, and we note these level-2 links can form triangles with seed nodes and common-neighbor nodes. It inspires us that triangle information is very useful in estimating similarities between nodes. It is notable that a recently proposed triangle growing network model can generate networks with various key features that can be observed in most real-world data sets, with only two kinds of triangle growth mechanism \cite{wu2015Emergence}. That confirms the crucial importance of triangle information in link formation. Further, we find that there are also some other related triangle information can be considered, such as triangles formed by a common-neighbor and its neighbors. Therefore, we consider this kind of information in our new index, called CCLP, by employing clustering coefficient of common-neighbors, because clustering coefficient can reflect the density of triangles within a local network environment. The definition is given in equation (\ref{eq7}), in which $CC_z$ is defined as equation (\ref{eq8}).

\begin{equation}
CCLP(x,y) = \sum_{z\in\Gamma(x)\cap\Gamma(y)}CC_{z}
\label{eq7}
\end{equation}

\begin{equation}
CC_{z} = \frac{t_{z}}{k_{z}(k_{z}-1)/2}
\label{eq8}
\end{equation}
where $t_{z}$ is the number of triangles passing through node $z$  and $k_{z}$ is the degree of node $z$.

Figure \ref{fig:progress} shows the illustration of local similarity indices. From PA index to CN/JC, level-1 link information is proved more valuable in all one-step links. From AA/RA to CAR, level-2 link structure shows more power in improving the accuracy of link-prediction algorithms. In this paper, we extend the local link/triangle structure information to farther structures, i.e. links between other neighbors of common neighbors or other triangles passing through a common-neighbor. We also find an efficient way to convey this link-structure information by employing clustering coefficient of common-neighbor. The key advantage of employing clustering coefficient is that some important links that possess the same structural position with the candidate link can be conveyed.

\subsection{Estimation}
To estimate the predicted results comprehensively, here we employ three different estimators: precision, AUP and AUC. The basic preparation for calculation of the three estimators is the same. To test the precision of a prediction algorithm, the observed links $E$ is randomly divided into two parts: the training set $E_t$ is treated as known information, while the probe set $E_p$ is used for testing and no information in the probe set is allowed to be used for prediction. Obviously, $E = E_t\cup E_p$ and $E_t\cap E_p = null$. In this paper, we consider 10\% of links as test links. The accuracy results are averages of 1000 independent runs for small networks with less than 1000 nodes and averages of 300 runs for large networks with more than 1000 nodes.

Given the ranking of the non-observed links, the precision is defined as the ratio of relevant items selected to the number of items selected. That means if we take the top $L$ links as the predicted ones, among which $L_r$ links are right, then the precision can be defined as equation (\ref{eq9})\cite{herlocker2004evaluating}. In this paper, we choose $L$ as 20 for networks with less than 1000 links and 100 for networks with more than 1000 links. Higher precision indicates higher prediction accuracy. 

\begin{equation}
precision = \frac{L_{r}}{L}
\label{eq9}
\end{equation}

AUP is defined as the Area Under Precision curve, which is proposed by Cannistraci et al \cite{cannistraci2013link}. In this paper, the precision curve is achieved by using ten different values of $L$. The largest value of $L$ is defined as that in the above definition of precision and other values is an arithmetical sequence with the difference equal to one tenth of the largest $L$.  

AUC, the area under the receiver operating characteristic (ROC) curve, is a metric to quantify the accuracy of the prediction algorithms. In this situation, it can be interpreted as the probability that a randomly chosen missing link (belongs to $E_p$) is given a higher score than a randomly chosen nonexistent link (which belongs to $U-E$, where $U$ denotes the set of all node-pairs). In practice, the calculation of AUC is given as defined by equation (\ref{eq10}), where $n$ is the times of independent comparisons, $n'$ denotes the times the missing links having a higher score, and $n''$ counts situations they have the same score. A higher value of AUC indicates better results. Although there are some comments on AUC in link prediction estimation \cite{yang2014evaluating}, it is still a widly used index. Therefore, we also give the AUC results in this article.  

\begin{equation}
AUC = \frac{n'+0.5n''}{n}
\label{eq10}
\end{equation}

\subsection{Datasets}
In this article, we test link prediction algorithms on 10 real-world networks drawn from five various fields. Food \cite{ulanowicz2005network} and Grassland \cite{dawah1995structure} are two food web networks. Dolphins \cite{lusseau2003bottlenose} and Jazz \cite{gleiser2003community} are dolphins and musician social networks. Macaque \cite{kotter2004online} and Mouse \cite{bock2011network} are two neural networks. PB \cite{adamic2005political} and Email \cite{guimera2003self} are two social networks from electronic information systems. Grid \cite{watts1998collective} and INT \cite{spring2002measuring} are two artificial infrastructure networks. The basic topological features of these networks are given in Table \ref{tab:t1}, including number of nodes and edges, average degree, average shortest distance, clustering coefficient and $LCP_{-corr}$, which is defined in equation (\ref{eq11}).

\begin{table}[ht]
\centering
\begin{tabular}{|l|l|l|l|l|l|l|}
\hline
Nets        &N      &M     &$\langle k\rangle$     &$\langle d\rangle$     &CC    &$LCP_{-corr}$\\  
\hline 
Food&128&2075&32.42&1.78&0.31&0.91\\  
Grassland&75&113&3.01&3.88&0.34&0.42\\  
Dolphins&62&159&5.13&3.36&0.26&0.89\\  
Jazz&198&2742&27.7&2.24&0.62&0.95\\  
Macaque&94&1515&32.23&1.77&0.77&0.97\\  
Mouse&18&37&4.11&1.97&0.22&0.91\\  
PB&1222&16714&27.36&2.74&0.32&0.93\\  
Email&1133&5451&9.62&3.61&0.22&0.85\\  
Grid&4941&6594&2.67&15.87&0.08&0.78\\  
INT&5022&6258&2.49&6.0&0.01&0.81\\  
\hline
\end{tabular}
\caption{\label{tab:t1}The basic topological features of the 10 real-world networks. $N$ and $M$ are the total number of nodes and links, respectively. $\langle k\rangle$ is the average degree of the network. $\langle d\rangle$ is the average shortest distance between node pairs. CC is the clustering coefficient defined in equation (\ref{eq8}). $LCP_{-corr}$ is an index to estimate the correlation between CN (number of common-neighbors) and LCL (number of links between common-neighbors).}
\end{table}

\begin{equation}
LCP_{-corr} = \frac{cov(CN,LCL)}{\delta_{CN} \delta_{LCL}}, CN > 1,
\label{eq11}
\end{equation}
where $cov(X,Y)$ is the covariance of the variables $X$ and $Y$, and $\delta_{X}$ is the standard deviation of the variable $X$.

\section{Results}

\subsection{Accuracy}

\begin{table*}[ht]
\centering
\begin{tabular}{|l|l|l|l|l|l|l|l|}
\hline
precision&RP&CN&AA&RA&CAR&CCLP&CCLP/CAR\\ 
\hline 
Food&0.033 &0.087 &0.089 &0.086 &0.093 &\textbf{0.097} &+4.45\%\\ 
Grassland&0.004 &0.065 &\textbf{0.116} &0.115 &0.021 &0.102 &+386.19\%\\ 
Dolphins&0.009 &0.126 &0.119 &0.105 &\textbf{0.143} &0.112 &-27.91\%\\ 
Jazz&0.016 &0.823 &0.838 &0.821 &0.853 &\textbf{0.859} &+0.75\%\\ 
Macaque&0.051 &0.577 &0.578 &0.555 &\textbf{0.612} &0.608 &-0.66\%\\ 
Mouse&0.034 &0.045 &0.043 &0.043 &0.035 &\textbf{0.052} &+51.01\%\\ 
PB&0.002 &0.416 &0.377 &0.246 &\textbf{0.477} &0.404 &-18.04\%\\ 
Email&0.001 &0.285 &\textbf{0.314} &0.258 &0.305 &0.308 &+0.71\%\\ 
Grid&0.000 &0.125 &0.101 &0.081 &\textbf{0.178} &0.172 &-3.32\%\\ 
INT&0.000 &0.105 &0.107 &0.084 &0.081 &\textbf{0.172} &+111.05\%\\ 
\hline
\end{tabular}
\caption{\label{tab:t2}Link prediction accuracy of compared similarity indices and CCLP estimated by precision.}
\end{table*}

\begin{table*}[ht]
\centering
\begin{tabular}{|l|l|l|l|l|l|l|l|}
\hline
AUP&RP&CN&AA&RA&CAR&CCLP&CCLP/CAR\\  
\hline 
Food&0.038&0.112&0.114&0.106&\textbf{0.123}&\textbf{0.123}&0.00\%\\ 
Grassland&0.005&0.049&0.163&\textbf{0.167}&0.028&0.108&+285.71\%\\ 
Dolphins&0.009&\textbf{0.19}&0.164&0.147&0.181&0.144&-25.69\%\\ 
Jazz&0.016&0.888&0.894&0.884&0.927&\textbf{0.928}&+0.11\%\\ 
Macaque&0.05&0.647&0.645&0.603&\textbf{0.69}&0.678&-1.77\%\\ 
Mouse&0.033&0.047&0.048&0.047&0.07&\textbf{0.083}&+18.57\%\\ 
PB&0.002&0.459&0.393&0.16&\textbf{0.574}&0.477&-20.34\%\\ 
Email&0.001&0.351&0.387&0.316&0.416&\textbf{0.424}&+1.92\%\\ 
Grid&0&0.218&0.106&0.088&0.228&\textbf{0.248}&+8.77\%\\ 
INT&0&0.128&0.107&0.078&0.117&\textbf{0.218}&+86.32\%\\ 
\hline
\end{tabular}
\caption{\label{tab:t3}Link prediction accuracy of compared similarity indices and CCLP estimated by AUP.}
\end{table*}

Here we provide the results of accuracy for several common-neighbor-based similarity indices. CN, AA and RA have been shown to perform better than other classical common-neighbor-based similarity indices, so for simplicity we only choose CN, AA, RA and CAR to compare with CCLP index. Besides, we also give the results of a random predictor (RP).

The link prediction results estimated by precision are shown in Table \ref{tab:t2}. Highest scores among these indices are highlighted in boldface. Estimated in precision, CAR and CCLP perform better than other classical indices on the ten networks. They get four best results, respectively. For comparing CCLP and CAR more clearly, we also give a column of 'CCLP/CAR', which shows the ratio of improvement on each other. Positive value means CCLP performs better than CAR, and vise versa. It shows that CCLP index performs slightly better than CAR index with six better results. From the aspect of improvement ratio, CCLP can improve more than hundreds of percentages on Grassland and INT networks, while the largest improvement made by CAR on CCLP is up to 27.9\%. Since Grassland is not a typical LCP (Local community paradigm) network, which has a low $LCP_{-corr}$ value as shown in Talbe \ref{tab:t1}, CAR can not give precise predicting results. But CCLP does not suffer from this kind of limitation. 

Table \ref{tab:t3} presents the prediction results estimated by AUP. Still CCLP and CAR perform better than other classical indices. CCLP gets six best accuracy results and CAR gets three bests. At this time CCLP performs better than CAR index with six higher scores, as shown in the column of 'CCLP/CAR', and big improvements appear again on Grassland and INT networks.  

\begin{figure*}[ht]
\centering
\includegraphics[width=12cm]{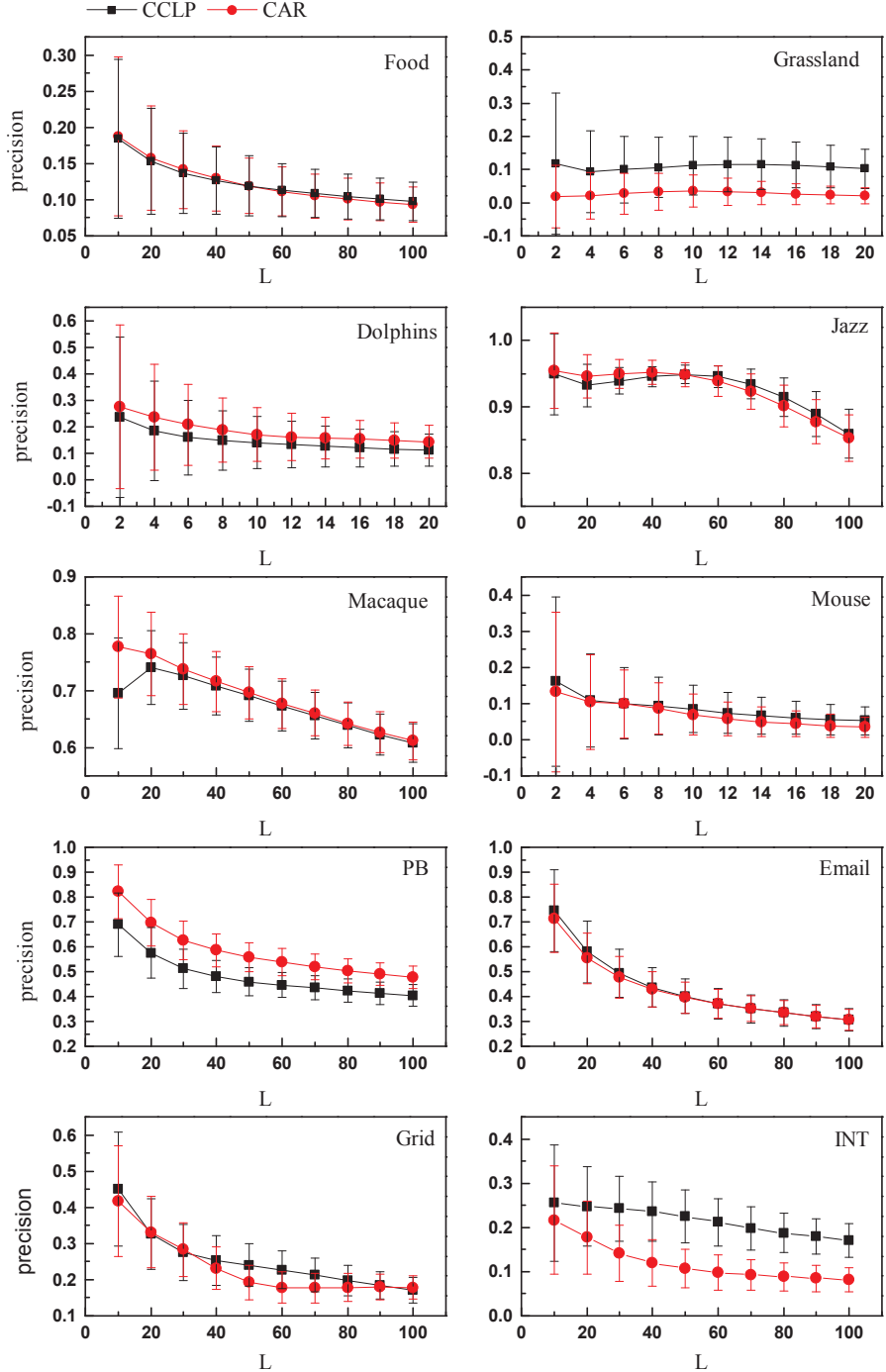}
\caption{Precision curve for CCLP and CAR indices on 10 tested networks.}
\label{fig:2}
\end{figure*} 

Further, we also demonstrate the precision curve of CCLP and CAR on the ten tested networks in Fig \ref{fig:2}. Different data points are generated corresponding to different parameter $L$ of precision estimator. Almost for all tested networks, precision decreases with the increase of $L$, except for Grassland network, in which the prediction results of CCLP and CAR are very stable despite of the variation of $L$. As a whole, it validates our proposition that increasing $L$ increases the difficulty of the prediction problem. 

From Fig \ref{fig:2}, it is clearly that CCLP performs better than CAR on Grassland, INT and Grid networks and CAR gets better results on PB and Dolphins networks. As the authors of CAR index analyzed, CAR is only good at predicting missing link for typical LCP networks with high $LCP_{-corr}$. Most networks on which CCLP can perform better than CAR have relatively small $LCP_{-corr}$. It indicates that CCLP index is a good complement of CAR index when facing atypical LCP (Local community paradigm) networks.

In our experiments, we also observe another interesting phenomenon. On Food networks, classical similarity indices perform even worse than the random predictor (RP) estimated by both precision and AUP. By punishing large degree nodes, AA and RA can perform better than CN, but their prediction results are still worse than the random predictor. With the similar form of definition with AA and RA, CCLP estimates the contributions of common neighbors with clustering coefficient and achieves useful structural information on Food network.

\begin{table*}[ht]
\centering
\begin{tabular}{|l|l|l|l|l|l|l|l|}
\hline
AUC&RP&CN&AA&RA&CAR&CCLP&CCLP/CAR\\  
\hline 
Food&0.500 &0.608 &0.611 &0.614 &0.621 &\textbf{0.636} &+2.29\%\\ 
Grassland&0.500 &0.784 &\textbf{0.797} &\textbf{0.797} &0.516 &0.790 &+53.08\%\\ 
Dolphins&0.500 &0.802 &\textbf{0.805} &\textbf{0.805} &0.647 &0.802 &+23.91\%\\ 
Jazz&0.500 &0.956 &0.963 &\textbf{0.972} &0.955 &0.960 &+0.59\%\\ 
Macaque&0.500 &0.945 &0.947 &\textbf{0.949} &0.945 &0.948 &+0.32\%\\ 
Mouse&0.500 &0.478 &0.482 &0.482 &\textbf{0.567} &0.499 &-13.50\%\\ 
PB&0.500 &0.924 &0.927 &\textbf{0.929} &0.896 &0.927 &+3.39\%\\ 
Email&0.500 &0.856 &\textbf{0.858} &\textbf{0.858} &0.703 &0.857 &+21.95\%\\ 
Grid&0.500 &\textbf{0.626} &\textbf{0.626} &\textbf{0.626} &0.517 &\textbf{0.626} &+21.09\%\\ 
INT&0.500 &0.652 &0.652 &0.652 &0.529 &\textbf{0.653} &+23.42\%\\ 
\hline
\end{tabular}
\caption{\label{tab:t4}Link prediction accuracy of compared similarity indices and CCLP estimated by AUC.}
\end{table*}

Table \ref{tab:t4} gives the accuracy results estimated by AUC. AUC estimator evaluates the accuracy considering the whole ranking results. RA index performs the bests in Table \ref{tab:t4}, and the differences among results of RA, AA and CCLP are very small. Comparing with CAR index, CCLP obviously performs better estimated by AUC.

\subsection{Efficiency} 

In this section, we will validate the efficiency of CCLP index. Before assigning similarity scores for all pairs of nodes, CCLP needs $O(N)$ times to calculate clustering coefficient for each node, while CAR needs $O(N^2)$ times to calculate LCL for each pair of nodes. Thus CCLP costs a bit less computational time than CAR index. The average complexity of assigning final similarity scores is $O(N^2\langle k\rangle^2)$, where $\langle k\rangle$ is the average degree of a network. That is the same for most common-neighbor-based similarity indices, even for the simplest index, CN. The results shown in Table \ref{tab:t5} validate the above analysis. When implementing in the same way (i.e. in loop-based matlab implementation), the time costs don't differ too much. CAR index is the most complex index, but only costs a bit more time than other indices. Since other indices can be implemented in matrix operation, we also give the corresponding time costs in Table \ref{tab:t6}.

\begin{table}[ht]
\centering
\begin{tabular}{|l|l|l|l|l|l|}
\hline
time&CN&AA&RA&CAR&CCLP\\ 
\hline 
Food&1.6041 &1.6255 &1.6134 &1.6618 &1.6173 \\ 
Grassland&0.6772 &0.6984 &0.6929 &0.7071 &0.6933 \\ 
Dolphins&0.4370 &0.4420 &0.4387 &0.4495 &0.4398 \\ 
Jazz&4.3844 &4.4377 &4.4052 &4.5210 &4.4136 \\ 
Macaque&0.7811 &0.7913 &0.7854 &0.7878 &0.8078 \\ 
Mouse&0.0338 &0.0342 &0.0339 &0.0347 &0.0342 \\ 
PB&188.46 &190.61 &189.26 &193.93 &189.48 \\ 
Email&164.32 &169.35 &168.09 &171.23 &168.02 \\ 
Grid&2985.13 &3022.45 &2999.78 &3071.21 &3001.89 \\ 
INT&3008.31 &3045.30 &3021.85 &3097.72 &3024.49 \\ 
\hline
\end{tabular}
\caption{\label{tab:t5}Computation time (in second) of similarity matrices for compared indices on 10 tested networks implementing in loop-structure.}
\end{table}

\begin{table}[ht]
\centering
\begin{tabular}{|l|l|l|l|l|}
\hline
time&CN&AA&RA&CCLP\\ 
\hline 
Food&0.0037 &0.0019 &0.0017 &0.0038 \\ 
Grassland&0.0002 &0.0003 &0.0002 &0.0003 \\ 
Dolphins&0.0002 &0.0003 &0.0002 &0.0003 \\ 
Jazz&0.0024 &0.0029 &0.0026 &0.0057 \\ 
Macaque&0.0009 &0.0011 &0.0010 &0.0019 \\ 
Mouse&0.0001 &0.0002 &0.0001 &0.0002 \\ 
PB&0.0533 &0.0698 &0.0696 &0.1241 \\ 
Email&0.0158 &0.0313 &0.0310 &0.0466 \\ 
Grid&0.1829 &0.4110 &0.4438 &0.5167 \\ 
INT&0.1811 &0.4060 &0.4350 &0.3451 \\ 
\hline
\end{tabular}
\caption{\label{tab:t6}Computation time (in second) of similarity matrices for compared indices except CAR index on 10 tested networks implementing in matrix operation.}
\end{table}

\section{CONCLUSION}

In recent years, some sophisticated link prediction models and methods, such as HSM \cite{clauset2008hierarchical}, SBM \cite{guimera2009missing}, and SPM \cite{lu2015toward}, have been proposed. Not only these methods can give very good prediction results in some networks, another significance of these methods is to give insights into the mechanism of link formation, network evolution, and even the link predictability \cite{lu2015toward}. Although the study on sophisticated models is undoubtedly significant, we think that at least currently efficient link prediction methods should not be neglected until the computational complexity is not a problem for sophisticated technics.

In this paper we exploit the power of more local link/triangle structure information, which can be conveyed by clustering coefficient of common neighbors efficiently. CAR index employs local link structure information, called LCL, which belongs to a pair of nodes. It is very helpful in improving link prediction accuracy, yet not suitable for networks with not very high $LCP_{-corr}$. Experimental results on real-world networks demonstrate that with clustering coefficient CCLP index can perform comparatively with CAR index as a whole. For networks with not very high $LCP_{-corr}$, CCLP index is a good complement to CAR index with better link prediction accuracy. 

The similarity of CCLP and CAR index is that both of them employ some local link/triangle structure information rather than simple degree information. The difference is that CCLP utilizes more link/triangle structural information, while the information used by CAR is more specific to a pair of nodes. LCL belongs to a pair of candidate seed nodes, but clustering coefficient of a common-neighbor is the same for all different pairs of nodes. That may be the reason why CCLP can not perform better than CAR index significantly. Certainly, to attain more specific information CAR is only sutitable for networks with high $LCP_{-corr}$. In the future work, considering both of the two factors in one index at the same time may be a good direction to design better Node-Neighborhood-based similarity indices.

\begin{acknowledgments}
 This work is supported by the National Natural Science Foundation of China (Grants No. 61403023) and China Postdoctoral Science Foundation (Grants No. 2014M550604).
\end{acknowledgments}

\bibliography{CCLPphysicaA}

\begin{thebibliography}{47}%
\makeatletter
\providecommand \@ifxundefined [1]{%
 \@ifx{#1\undefined}
}%
\providecommand \@ifnum [1]{%
 \ifnum #1\expandafter \@firstoftwo
 \else \expandafter \@secondoftwo
 \fi
}%
\providecommand \@ifx [1]{%
 \ifx #1\expandafter \@firstoftwo
 \else \expandafter \@secondoftwo
 \fi
}%
\providecommand \natexlab [1]{#1}%
\providecommand \enquote  [1]{``#1''}%
\providecommand \bibnamefont  [1]{#1}%
\providecommand \bibfnamefont [1]{#1}%
\providecommand \citenamefont [1]{#1}%
\providecommand \href@noop [0]{\@secondoftwo}%
\providecommand \href [0]{\begingroup \@sanitize@url \@href}%
\providecommand \@href[1]{\@@startlink{#1}\@@href}%
\providecommand \@@href[1]{\endgroup#1\@@endlink}%
\providecommand \@sanitize@url [0]{\catcode `\\12\catcode `\$12\catcode
  `\&12\catcode `\#12\catcode `\^12\catcode `\_12\catcode `\%12\relax}%
\providecommand \@@startlink[1]{}%
\providecommand \@@endlink[0]{}%
\providecommand \url  [0]{\begingroup\@sanitize@url \@url }%
\providecommand \@url [1]{\endgroup\@href {#1}{\urlprefix }}%
\providecommand \urlprefix  [0]{URL }%
\providecommand \Eprint [0]{\href }%
\providecommand \doibase [0]{http://dx.doi.org/}%
\providecommand \selectlanguage [0]{\@gobble}%
\providecommand \bibinfo  [0]{\@secondoftwo}%
\providecommand \bibfield  [0]{\@secondoftwo}%
\providecommand \translation [1]{[#1]}%
\providecommand \BibitemOpen [0]{}%
\providecommand \bibitemStop [0]{}%
\providecommand \bibitemNoStop [0]{.\EOS\space}%
\providecommand \EOS [0]{\spacefactor3000\relax}%
\providecommand \BibitemShut  [1]{\csname bibitem#1\endcsname}%
\let\auto@bib@innerbib\@empty
\bibitem [{\citenamefont {Albert}\ and\ \citenamefont
  {Barab{\'a}si}(2002)}]{albert2002statistical}%
  \BibitemOpen
  \bibfield  {author} {\bibinfo {author} {\bibfnamefont {R.}~\bibnamefont
  {Albert}}\ and\ \bibinfo {author} {\bibfnamefont {A.-L.}\ \bibnamefont
  {Barab{\'a}si}},\ }\href@noop {} {\bibfield  {journal} {\bibinfo  {journal}
  {Reviews of modern physics}\ }\textbf {\bibinfo {volume} {74}},\ \bibinfo
  {pages} {47} (\bibinfo {year} {2002})}\BibitemShut {NoStop}%
\bibitem [{\citenamefont {Boccaletti}\ \emph {et~al.}(2006)\citenamefont
  {Boccaletti}, \citenamefont {Latora}, \citenamefont {Moreno}, \citenamefont
  {Chavez},\ and\ \citenamefont {Hwang}}]{boccaletti2006complex}%
  \BibitemOpen
  \bibfield  {author} {\bibinfo {author} {\bibfnamefont {S.}~\bibnamefont
  {Boccaletti}}, \bibinfo {author} {\bibfnamefont {V.}~\bibnamefont {Latora}},
  \bibinfo {author} {\bibfnamefont {Y.}~\bibnamefont {Moreno}}, \bibinfo
  {author} {\bibfnamefont {M.}~\bibnamefont {Chavez}}, \ and\ \bibinfo {author}
  {\bibfnamefont {D.-U.}\ \bibnamefont {Hwang}},\ }\href@noop {} {\bibfield
  {journal} {\bibinfo  {journal} {Physics Reports}\ }\textbf {\bibinfo {volume}
  {424}},\ \bibinfo {pages} {175} (\bibinfo {year} {2006})}\BibitemShut
  {NoStop}%
\bibitem [{\citenamefont {Costa}\ \emph {et~al.}(2007)\citenamefont {Costa},
  \citenamefont {Rodrigues}, \citenamefont {Travieso},\ and\ \citenamefont
  {Villas~Boas}}]{costa2007characterization}%
  \BibitemOpen
  \bibfield  {author} {\bibinfo {author} {\bibfnamefont {L.~d.~F.}\
  \bibnamefont {Costa}}, \bibinfo {author} {\bibfnamefont {F.~A.}\ \bibnamefont
  {Rodrigues}}, \bibinfo {author} {\bibfnamefont {G.}~\bibnamefont {Travieso}},
  \ and\ \bibinfo {author} {\bibfnamefont {P.~R.}\ \bibnamefont
  {Villas~Boas}},\ }\href@noop {} {\bibfield  {journal} {\bibinfo  {journal}
  {Advances in Physics}\ }\textbf {\bibinfo {volume} {56}},\ \bibinfo {pages}
  {167} (\bibinfo {year} {2007})}\BibitemShut {NoStop}%
\bibitem [{\citenamefont {Bianconi}\ \emph {et~al.}(2009)\citenamefont
  {Bianconi}, \citenamefont {Pin},\ and\ \citenamefont
  {Marsili}}]{bianconi2009assessing}%
  \BibitemOpen
  \bibfield  {author} {\bibinfo {author} {\bibfnamefont {G.}~\bibnamefont
  {Bianconi}}, \bibinfo {author} {\bibfnamefont {P.}~\bibnamefont {Pin}}, \
  and\ \bibinfo {author} {\bibfnamefont {M.}~\bibnamefont {Marsili}},\
  }\href@noop {} {\bibfield  {journal} {\bibinfo  {journal} {Proceedings of the
  National Academy of Sciences}\ }\textbf {\bibinfo {volume} {106}},\ \bibinfo
  {pages} {11433} (\bibinfo {year} {2009})}\BibitemShut {NoStop}%
\bibitem [{\citenamefont {Shen}\ and\ \citenamefont
  {Barab{\'a}si}(2014)}]{shen2014collective}%
  \BibitemOpen
  \bibfield  {author} {\bibinfo {author} {\bibfnamefont {H.-W.}\ \bibnamefont
  {Shen}}\ and\ \bibinfo {author} {\bibfnamefont {A.-L.}\ \bibnamefont
  {Barab{\'a}si}},\ }\href@noop {} {\bibfield  {journal} {\bibinfo  {journal}
  {Proceedings of the National Academy of Sciences}\ }\textbf {\bibinfo
  {volume} {111}},\ \bibinfo {pages} {12325} (\bibinfo {year}
  {2014})}\BibitemShut {NoStop}%
\bibitem [{\citenamefont {Holme}\ and\ \citenamefont
  {Huss}(2005)}]{holme2005role}%
  \BibitemOpen
  \bibfield  {author} {\bibinfo {author} {\bibfnamefont {P.}~\bibnamefont
  {Holme}}\ and\ \bibinfo {author} {\bibfnamefont {M.}~\bibnamefont {Huss}},\
  }\href@noop {} {\bibfield  {journal} {\bibinfo  {journal} {Journal of The
  Royal Society Interface}\ }\textbf {\bibinfo {volume} {2}},\ \bibinfo {pages}
  {327} (\bibinfo {year} {2005})}\BibitemShut {NoStop}%
\bibitem [{\citenamefont {Gallagher}\ \emph {et~al.}(2008)\citenamefont
  {Gallagher}, \citenamefont {Tong}, \citenamefont {Eliassi-Rad},\ and\
  \citenamefont {Faloutsos}}]{gallagher2008using}%
  \BibitemOpen
  \bibfield  {author} {\bibinfo {author} {\bibfnamefont {B.}~\bibnamefont
  {Gallagher}}, \bibinfo {author} {\bibfnamefont {H.}~\bibnamefont {Tong}},
  \bibinfo {author} {\bibfnamefont {T.}~\bibnamefont {Eliassi-Rad}}, \ and\
  \bibinfo {author} {\bibfnamefont {C.}~\bibnamefont {Faloutsos}},\ }in\
  \href@noop {} {\emph {\bibinfo {booktitle} {Proceedings of the 14th ACM
  SIGKDD international conference on Knowledge discovery and data mining}}}\
  (\bibinfo {organization} {ACM},\ \bibinfo {year} {2008})\ pp.\ \bibinfo
  {pages} {256--264}\BibitemShut {NoStop}%
\bibitem [{\citenamefont {Wang}\ \emph {et~al.}(2012)\citenamefont {Wang},
  \citenamefont {Zhang},\ and\ \citenamefont {Zhou}}]{wang2012evaluating}%
  \BibitemOpen
  \bibfield  {author} {\bibinfo {author} {\bibfnamefont {W.-Q.}\ \bibnamefont
  {Wang}}, \bibinfo {author} {\bibfnamefont {Q.-M.}\ \bibnamefont {Zhang}}, \
  and\ \bibinfo {author} {\bibfnamefont {T.}~\bibnamefont {Zhou}},\ }\href@noop
  {} {\bibfield  {journal} {\bibinfo  {journal} {EPL (Europhysics Letters)}\
  }\textbf {\bibinfo {volume} {98}},\ \bibinfo {pages} {28004} (\bibinfo {year}
  {2012})}\BibitemShut {NoStop}%
\bibitem [{\citenamefont {Zhang}\ \emph {et~al.}(2015)\citenamefont {Zhang},
  \citenamefont {Xu}, \citenamefont {Zhu},\ and\ \citenamefont
  {Zhou}}]{zhang2015Measuring}%
  \BibitemOpen
  \bibfield  {author} {\bibinfo {author} {\bibfnamefont {Q.-M.}\ \bibnamefont
  {Zhang}}, \bibinfo {author} {\bibfnamefont {X.-K.}\ \bibnamefont {Xu}},
  \bibinfo {author} {\bibfnamefont {Y.-X.}\ \bibnamefont {Zhu}}, \ and\
  \bibinfo {author} {\bibfnamefont {T.}~\bibnamefont {Zhou}},\ }\href@noop {}
  {\bibfield  {journal} {\bibinfo  {journal} {Scientific reports}\ }\textbf
  {\bibinfo {volume} {5}},\ \bibinfo {pages} {10350} (\bibinfo {year}
  {2015})}\BibitemShut {NoStop}%
\bibitem [{\citenamefont {Liben-Nowell}\ and\ \citenamefont
  {Kleinberg}(2007)}]{liben2007link}%
  \BibitemOpen
  \bibfield  {author} {\bibinfo {author} {\bibfnamefont {D.}~\bibnamefont
  {Liben-Nowell}}\ and\ \bibinfo {author} {\bibfnamefont {J.}~\bibnamefont
  {Kleinberg}},\ }\href@noop {} {\bibfield  {journal} {\bibinfo  {journal}
  {Journal of the American society for information science and technology}\
  }\textbf {\bibinfo {volume} {58}},\ \bibinfo {pages} {1019} (\bibinfo {year}
  {2007})}\BibitemShut {NoStop}%
\bibitem [{\citenamefont {L{\"u}}\ and\ \citenamefont
  {Zhou}(2011)}]{lu2011link}%
  \BibitemOpen
  \bibfield  {author} {\bibinfo {author} {\bibfnamefont {L.}~\bibnamefont
  {L{\"u}}}\ and\ \bibinfo {author} {\bibfnamefont {T.}~\bibnamefont {Zhou}},\
  }\href@noop {} {\bibfield  {journal} {\bibinfo  {journal} {Physica A:
  Statistical Mechanics and its Applications}\ }\textbf {\bibinfo {volume}
  {390}},\ \bibinfo {pages} {1150} (\bibinfo {year} {2011})}\BibitemShut
  {NoStop}%
\bibitem [{\citenamefont {Newman}(2001)}]{newman2001clustering}%
  \BibitemOpen
  \bibfield  {author} {\bibinfo {author} {\bibfnamefont {M.~E.}\ \bibnamefont
  {Newman}},\ }\href@noop {} {\bibfield  {journal} {\bibinfo  {journal}
  {Physical Review E}\ }\textbf {\bibinfo {volume} {64}},\ \bibinfo {pages}
  {025102} (\bibinfo {year} {2001})}\BibitemShut {NoStop}%
\bibitem [{\citenamefont {Lorrain}\ and\ \citenamefont
  {White}(1971)}]{lorrain1971structural}%
  \BibitemOpen
  \bibfield  {author} {\bibinfo {author} {\bibfnamefont {F.}~\bibnamefont
  {Lorrain}}\ and\ \bibinfo {author} {\bibfnamefont {H.~C.}\ \bibnamefont
  {White}},\ }\href@noop {} {\bibfield  {journal} {\bibinfo  {journal} {The
  Journal of mathematical sociology}\ }\textbf {\bibinfo {volume} {1}},\
  \bibinfo {pages} {49} (\bibinfo {year} {1971})}\BibitemShut {NoStop}%
\bibitem [{\citenamefont {Jaccard}(1901)}]{jaccard1901distribution}%
  \BibitemOpen
  \bibfield  {author} {\bibinfo {author} {\bibfnamefont {P.}~\bibnamefont
  {Jaccard}},\ }\href@noop {} {\bibfield  {journal} {\bibinfo  {journal}
  {Bulletin de la Societe Vaudoise des Sciences Naturelles}\ }\textbf {\bibinfo
  {volume} {37}},\ \bibinfo {pages} {241} (\bibinfo {year} {1901})}\BibitemShut
  {NoStop}%
\bibitem [{\citenamefont {Adamic}\ and\ \citenamefont
  {Adar}(2003)}]{adamic2003friends}%
  \BibitemOpen
  \bibfield  {author} {\bibinfo {author} {\bibfnamefont {L.~A.}\ \bibnamefont
  {Adamic}}\ and\ \bibinfo {author} {\bibfnamefont {E.}~\bibnamefont {Adar}},\
  }\href@noop {} {\bibfield  {journal} {\bibinfo  {journal} {Social Networks}\
  }\textbf {\bibinfo {volume} {25}},\ \bibinfo {pages} {211} (\bibinfo {year}
  {2003})}\BibitemShut {NoStop}%
\bibitem [{\citenamefont {Zhou}\ \emph {et~al.}(2009)\citenamefont {Zhou},
  \citenamefont {L{\"u}},\ and\ \citenamefont {Zhang}}]{zhou2009predicting}%
  \BibitemOpen
  \bibfield  {author} {\bibinfo {author} {\bibfnamefont {T.}~\bibnamefont
  {Zhou}}, \bibinfo {author} {\bibfnamefont {L.}~\bibnamefont {L{\"u}}}, \ and\
  \bibinfo {author} {\bibfnamefont {Y.-C.}\ \bibnamefont {Zhang}},\ }\href@noop
  {} {\bibfield  {journal} {\bibinfo  {journal} {The European Physical Journal
  B-Condensed Matter and Complex Systems}\ }\textbf {\bibinfo {volume} {71}},\
  \bibinfo {pages} {623} (\bibinfo {year} {2009})}\BibitemShut {NoStop}%
\bibitem [{\citenamefont {Cannistraci}\ \emph {et~al.}(2013)\citenamefont
  {Cannistraci}, \citenamefont {Alanis-Lobato},\ and\ \citenamefont
  {Ravasi}}]{cannistraci2013link}%
  \BibitemOpen
  \bibfield  {author} {\bibinfo {author} {\bibfnamefont {C.~V.}\ \bibnamefont
  {Cannistraci}}, \bibinfo {author} {\bibfnamefont {G.}~\bibnamefont
  {Alanis-Lobato}}, \ and\ \bibinfo {author} {\bibfnamefont {T.}~\bibnamefont
  {Ravasi}},\ }\href@noop {} {\bibfield  {journal} {\bibinfo  {journal}
  {Scientific reports}\ }\textbf {\bibinfo {volume} {3}},\ \bibinfo {pages}
  {1613} (\bibinfo {year} {2013})}\BibitemShut {NoStop}%
\bibitem [{\citenamefont {Katz}(1953)}]{katz1953new}%
  \BibitemOpen
  \bibfield  {author} {\bibinfo {author} {\bibfnamefont {L.}~\bibnamefont
  {Katz}},\ }\href@noop {} {\bibfield  {journal} {\bibinfo  {journal}
  {Psychometrika}\ }\textbf {\bibinfo {volume} {18}},\ \bibinfo {pages} {39}
  (\bibinfo {year} {1953})}\BibitemShut {NoStop}%
\bibitem [{\citenamefont {Jeh}\ and\ \citenamefont
  {Widom}(2002)}]{jeh2002simrank}%
  \BibitemOpen
  \bibfield  {author} {\bibinfo {author} {\bibfnamefont {G.}~\bibnamefont
  {Jeh}}\ and\ \bibinfo {author} {\bibfnamefont {J.}~\bibnamefont {Widom}},\
  }in\ \href@noop {} {\emph {\bibinfo {booktitle} {Proceedings of the eighth
  ACM SIGKDD international conference on Knowledge discovery and data
  mining}}}\ (\bibinfo {organization} {ACM},\ \bibinfo {year} {2002})\ pp.\
  \bibinfo {pages} {538--543}\BibitemShut {NoStop}%
\bibitem [{\citenamefont {G{\"o}bel}\ and\ \citenamefont
  {Jagers}(1974)}]{gobel1974random}%
  \BibitemOpen
  \bibfield  {author} {\bibinfo {author} {\bibfnamefont {F.}~\bibnamefont
  {G{\"o}bel}}\ and\ \bibinfo {author} {\bibfnamefont {A.}~\bibnamefont
  {Jagers}},\ }\href@noop {} {\bibfield  {journal} {\bibinfo  {journal}
  {Stochastic processes and their applications}\ }\textbf {\bibinfo {volume}
  {2}},\ \bibinfo {pages} {311} (\bibinfo {year} {1974})}\BibitemShut {NoStop}%
\bibitem [{\citenamefont {Fouss}\ \emph {et~al.}(2007)\citenamefont {Fouss},
  \citenamefont {Pirotte}, \citenamefont {Renders},\ and\ \citenamefont
  {Saerens}}]{fouss2007random}%
  \BibitemOpen
  \bibfield  {author} {\bibinfo {author} {\bibfnamefont {F.}~\bibnamefont
  {Fouss}}, \bibinfo {author} {\bibfnamefont {A.}~\bibnamefont {Pirotte}},
  \bibinfo {author} {\bibfnamefont {J.-M.}\ \bibnamefont {Renders}}, \ and\
  \bibinfo {author} {\bibfnamefont {M.}~\bibnamefont {Saerens}},\ }\href@noop
  {} {\bibfield  {journal} {\bibinfo  {journal} {Knowledge and data
  engineering, ieee transactions on}\ }\textbf {\bibinfo {volume} {19}},\
  \bibinfo {pages} {355} (\bibinfo {year} {2007})}\BibitemShut {NoStop}%
\bibitem [{\citenamefont {L{\"u}}\ \emph {et~al.}(2009)\citenamefont {L{\"u}},
  \citenamefont {Jin},\ and\ \citenamefont {Zhou}}]{lu2009similarity}%
  \BibitemOpen
  \bibfield  {author} {\bibinfo {author} {\bibfnamefont {L.}~\bibnamefont
  {L{\"u}}}, \bibinfo {author} {\bibfnamefont {C.-H.}\ \bibnamefont {Jin}}, \
  and\ \bibinfo {author} {\bibfnamefont {T.}~\bibnamefont {Zhou}},\ }\href@noop
  {} {\bibfield  {journal} {\bibinfo  {journal} {Physical Review E}\ }\textbf
  {\bibinfo {volume} {80}},\ \bibinfo {pages} {046122} (\bibinfo {year}
  {2009})}\BibitemShut {NoStop}%
\bibitem [{\citenamefont {Sun}\ \emph {et~al.}(2009)\citenamefont {Sun},
  \citenamefont {Zhou}, \citenamefont {Liu}, \citenamefont {Liu}, \citenamefont
  {Jia},\ and\ \citenamefont {Wang}}]{sun2009information}%
  \BibitemOpen
  \bibfield  {author} {\bibinfo {author} {\bibfnamefont {D.}~\bibnamefont
  {Sun}}, \bibinfo {author} {\bibfnamefont {T.}~\bibnamefont {Zhou}}, \bibinfo
  {author} {\bibfnamefont {J.-G.}\ \bibnamefont {Liu}}, \bibinfo {author}
  {\bibfnamefont {R.-R.}\ \bibnamefont {Liu}}, \bibinfo {author} {\bibfnamefont
  {C.-X.}\ \bibnamefont {Jia}}, \ and\ \bibinfo {author} {\bibfnamefont
  {B.-H.}\ \bibnamefont {Wang}},\ }\href@noop {} {\bibfield  {journal}
  {\bibinfo  {journal} {Physical Review E}\ }\textbf {\bibinfo {volume} {80}},\
  \bibinfo {pages} {017101} (\bibinfo {year} {2009})}\BibitemShut {NoStop}%
\bibitem [{\citenamefont {Chebotarev}\ and\ \citenamefont
  {Shamis}(1997)}]{chebotarev1997matrix}%
  \BibitemOpen
  \bibfield  {author} {\bibinfo {author} {\bibfnamefont {P.~Y.}\ \bibnamefont
  {Chebotarev}}\ and\ \bibinfo {author} {\bibfnamefont {E.}~\bibnamefont
  {Shamis}},\ }\href@noop {} {\bibfield  {journal} {\bibinfo  {journal}
  {Avtomatika i Telemekhanika}\ ,\ \bibinfo {pages} {125}} (\bibinfo {year}
  {1997})}\BibitemShut {NoStop}%
\bibitem [{\citenamefont {Clauset}\ \emph {et~al.}(2008)\citenamefont
  {Clauset}, \citenamefont {Moore},\ and\ \citenamefont
  {Newman}}]{clauset2008hierarchical}%
  \BibitemOpen
  \bibfield  {author} {\bibinfo {author} {\bibfnamefont {A.}~\bibnamefont
  {Clauset}}, \bibinfo {author} {\bibfnamefont {C.}~\bibnamefont {Moore}}, \
  and\ \bibinfo {author} {\bibfnamefont {M.~E.}\ \bibnamefont {Newman}},\
  }\href@noop {} {\bibfield  {journal} {\bibinfo  {journal} {Nature}\ }\textbf
  {\bibinfo {volume} {453}},\ \bibinfo {pages} {98} (\bibinfo {year}
  {2008})}\BibitemShut {NoStop}%
\bibitem [{\citenamefont {Redner}(2008)}]{redner2008networks}%
  \BibitemOpen
  \bibfield  {author} {\bibinfo {author} {\bibfnamefont {S.}~\bibnamefont
  {Redner}},\ }\href@noop {} {\bibfield  {journal} {\bibinfo  {journal}
  {Nature}\ }\textbf {\bibinfo {volume} {453}},\ \bibinfo {pages} {47}
  (\bibinfo {year} {2008})}\BibitemShut {NoStop}%
\bibitem [{\citenamefont {Guimer{\`a}}\ and\ \citenamefont
  {Sales-Pardo}(2009)}]{guimera2009missing}%
  \BibitemOpen
  \bibfield  {author} {\bibinfo {author} {\bibfnamefont {R.}~\bibnamefont
  {Guimer{\`a}}}\ and\ \bibinfo {author} {\bibfnamefont {M.}~\bibnamefont
  {Sales-Pardo}},\ }\href@noop {} {\bibfield  {journal} {\bibinfo  {journal}
  {Proceedings of the National Academy of Sciences}\ }\textbf {\bibinfo
  {volume} {106}},\ \bibinfo {pages} {22073} (\bibinfo {year}
  {2009})}\BibitemShut {NoStop}%
\bibitem [{\citenamefont {L{\"u}}\ \emph {et~al.}(2015)\citenamefont {L{\"u}},
  \citenamefont {Pan}, \citenamefont {Zhou}, \citenamefont {Zhang},\ and\
  \citenamefont {Stanley}}]{lu2015toward}%
  \BibitemOpen
  \bibfield  {author} {\bibinfo {author} {\bibfnamefont {L.}~\bibnamefont
  {L{\"u}}}, \bibinfo {author} {\bibfnamefont {L.}~\bibnamefont {Pan}},
  \bibinfo {author} {\bibfnamefont {T.}~\bibnamefont {Zhou}}, \bibinfo {author}
  {\bibfnamefont {Y.-C.}\ \bibnamefont {Zhang}}, \ and\ \bibinfo {author}
  {\bibfnamefont {H.~E.}\ \bibnamefont {Stanley}},\ }\href@noop {} {\bibfield
  {journal} {\bibinfo  {journal} {Proceedings of the National Academy of
  Sciences}\ }\textbf {\bibinfo {volume} {112}},\ \bibinfo {pages} {2325}
  (\bibinfo {year} {2015})}\BibitemShut {NoStop}%
\bibitem [{\citenamefont {Liu}\ \emph {et~al.}(2013)\citenamefont {Liu},
  \citenamefont {He}, \citenamefont {Kapoor},\ and\ \citenamefont
  {Srivastava}}]{liu2013correlations}%
  \BibitemOpen
  \bibfield  {author} {\bibinfo {author} {\bibfnamefont {Z.}~\bibnamefont
  {Liu}}, \bibinfo {author} {\bibfnamefont {J.-L.}\ \bibnamefont {He}},
  \bibinfo {author} {\bibfnamefont {K.}~\bibnamefont {Kapoor}}, \ and\ \bibinfo
  {author} {\bibfnamefont {J.}~\bibnamefont {Srivastava}},\ }\href@noop {}
  {\bibfield  {journal} {\bibinfo  {journal} {PloS ONE}\ }\textbf {\bibinfo
  {volume} {8}},\ \bibinfo {pages} {e72908} (\bibinfo {year}
  {2013})}\BibitemShut {NoStop}%
\bibitem [{\citenamefont {Yan}\ and\ \citenamefont
  {Gregory}(2012)}]{yan2012finding}%
  \BibitemOpen
  \bibfield  {author} {\bibinfo {author} {\bibfnamefont {B.}~\bibnamefont
  {Yan}}\ and\ \bibinfo {author} {\bibfnamefont {S.}~\bibnamefont {Gregory}},\
  }\href@noop {} {\bibfield  {journal} {\bibinfo  {journal} {Physical Review
  E}\ }\textbf {\bibinfo {volume} {85}},\ \bibinfo {pages} {056112} (\bibinfo
  {year} {2012})}\BibitemShut {NoStop}%
\bibitem [{\citenamefont {Soundarajan}\ and\ \citenamefont
  {Hopcroft}(2012)}]{soundarajan2012using}%
  \BibitemOpen
  \bibfield  {author} {\bibinfo {author} {\bibfnamefont {S.}~\bibnamefont
  {Soundarajan}}\ and\ \bibinfo {author} {\bibfnamefont {J.}~\bibnamefont
  {Hopcroft}},\ }in\ \href@noop {} {\emph {\bibinfo {booktitle} {Proceedings of
  the 21st international conference companion on World Wide Web}}}\ (\bibinfo
  {organization} {ACM},\ \bibinfo {year} {2012})\ pp.\ \bibinfo {pages}
  {607--608}\BibitemShut {NoStop}%
\bibitem [{\citenamefont {Feng}\ \emph {et~al.}(2012)\citenamefont {Feng},
  \citenamefont {Zhao},\ and\ \citenamefont {Xu}}]{feng2012link}%
  \BibitemOpen
  \bibfield  {author} {\bibinfo {author} {\bibfnamefont {X.}~\bibnamefont
  {Feng}}, \bibinfo {author} {\bibfnamefont {J.}~\bibnamefont {Zhao}}, \ and\
  \bibinfo {author} {\bibfnamefont {K.}~\bibnamefont {Xu}},\ }\href@noop {}
  {\bibfield  {journal} {\bibinfo  {journal} {The European Physical Journal B}\
  }\textbf {\bibinfo {volume} {85}},\ \bibinfo {pages} {1} (\bibinfo {year}
  {2012})}\BibitemShut {NoStop}%
\bibitem [{\citenamefont {Liu}\ \emph {et~al.}(2011)\citenamefont {Liu},
  \citenamefont {Zhang}, \citenamefont {L{\"u}},\ and\ \citenamefont
  {Zhou}}]{liu2011link}%
  \BibitemOpen
  \bibfield  {author} {\bibinfo {author} {\bibfnamefont {Z.}~\bibnamefont
  {Liu}}, \bibinfo {author} {\bibfnamefont {Q.-M.}\ \bibnamefont {Zhang}},
  \bibinfo {author} {\bibfnamefont {L.}~\bibnamefont {L{\"u}}}, \ and\ \bibinfo
  {author} {\bibfnamefont {T.}~\bibnamefont {Zhou}},\ }\href@noop {} {\bibfield
   {journal} {\bibinfo  {journal} {EPL (Europhysics Letters)}\ }\textbf
  {\bibinfo {volume} {96}},\ \bibinfo {pages} {48007} (\bibinfo {year}
  {2011})}\BibitemShut {NoStop}%
\bibitem [{\citenamefont {Wu}\ \emph {et~al.}(2015)\citenamefont {Wu},
  \citenamefont {Menichetti}, \citenamefont {Rahmede},\ and\ \citenamefont
  {Bianconi}}]{wu2015Emergence}%
  \BibitemOpen
  \bibfield  {author} {\bibinfo {author} {\bibfnamefont {Z.-H.}\ \bibnamefont
  {Wu}}, \bibinfo {author} {\bibfnamefont {G.}~\bibnamefont {Menichetti}},
  \bibinfo {author} {\bibfnamefont {C.}~\bibnamefont {Rahmede}}, \ and\
  \bibinfo {author} {\bibfnamefont {G.}~\bibnamefont {Bianconi}},\ }\href@noop
  {} {\bibfield  {journal} {\bibinfo  {journal} {Scientific reports}\ }\textbf
  {\bibinfo {volume} {5}},\ \bibinfo {pages} {10073} (\bibinfo {year}
  {2015})}\BibitemShut {NoStop}%
\bibitem [{\citenamefont {Herlocker}\ \emph {et~al.}(2004)\citenamefont
  {Herlocker}, \citenamefont {Konstan}, \citenamefont {Terveen},\ and\
  \citenamefont {Riedl}}]{herlocker2004evaluating}%
  \BibitemOpen
  \bibfield  {author} {\bibinfo {author} {\bibfnamefont {J.~L.}\ \bibnamefont
  {Herlocker}}, \bibinfo {author} {\bibfnamefont {J.~A.}\ \bibnamefont
  {Konstan}}, \bibinfo {author} {\bibfnamefont {L.~G.}\ \bibnamefont
  {Terveen}}, \ and\ \bibinfo {author} {\bibfnamefont {J.~T.}\ \bibnamefont
  {Riedl}},\ }\href@noop {} {\bibfield  {journal} {\bibinfo  {journal} {ACM
  Transactions on Information Systems (TOIS)}\ }\textbf {\bibinfo {volume}
  {22}},\ \bibinfo {pages} {5} (\bibinfo {year} {2004})}\BibitemShut {NoStop}%
\bibitem [{\citenamefont {Yang}\ \emph {et~al.}(2014)\citenamefont {Yang},
  \citenamefont {Lichtenwalter},\ and\ \citenamefont
  {Chawla}}]{yang2014evaluating}%
  \BibitemOpen
  \bibfield  {author} {\bibinfo {author} {\bibfnamefont {Y.}~\bibnamefont
  {Yang}}, \bibinfo {author} {\bibfnamefont {R.~N.}\ \bibnamefont
  {Lichtenwalter}}, \ and\ \bibinfo {author} {\bibfnamefont {N.~V.}\
  \bibnamefont {Chawla}},\ }\href@noop {} {\bibfield  {journal} {\bibinfo
  {journal} {Knowledge and Information Systems}\ ,\ \bibinfo {pages} {1}}
  (\bibinfo {year} {2014})}\BibitemShut {NoStop}%
\bibitem [{\citenamefont {Ulanowicz}\ and\ \citenamefont
  {DeAngelis}(2005)}]{ulanowicz2005network}%
  \BibitemOpen
  \bibfield  {author} {\bibinfo {author} {\bibfnamefont {R.~E.}\ \bibnamefont
  {Ulanowicz}}\ and\ \bibinfo {author} {\bibfnamefont {D.~L.}\ \bibnamefont
  {DeAngelis}},\ }\href@noop {} {\bibfield  {journal} {\bibinfo  {journal} {US
  Geological Survey Program on the South Florida Ecosystem}\ }\textbf {\bibinfo
  {volume} {114}} (\bibinfo {year} {2005})}\BibitemShut {NoStop}%
\bibitem [{\citenamefont {Dawah}\ \emph {et~al.}(1995)\citenamefont {Dawah},
  \citenamefont {Hawkins},\ and\ \citenamefont
  {Claridge}}]{dawah1995structure}%
  \BibitemOpen
  \bibfield  {author} {\bibinfo {author} {\bibfnamefont {H.~A.}\ \bibnamefont
  {Dawah}}, \bibinfo {author} {\bibfnamefont {B.~A.}\ \bibnamefont {Hawkins}},
  \ and\ \bibinfo {author} {\bibfnamefont {M.~F.}\ \bibnamefont {Claridge}},\
  }\href@noop {} {\bibfield  {journal} {\bibinfo  {journal} {Journal of animal
  ecology}\ }\textbf {\bibinfo {volume} {64}},\ \bibinfo {pages} {708}
  (\bibinfo {year} {1995})}\BibitemShut {NoStop}%
\bibitem [{\citenamefont {Lusseau}\ \emph {et~al.}(2003)\citenamefont
  {Lusseau}, \citenamefont {Schneider}, \citenamefont {Boisseau}, \citenamefont
  {Haase}, \citenamefont {Slooten},\ and\ \citenamefont
  {Dawson}}]{lusseau2003bottlenose}%
  \BibitemOpen
  \bibfield  {author} {\bibinfo {author} {\bibfnamefont {D.}~\bibnamefont
  {Lusseau}}, \bibinfo {author} {\bibfnamefont {K.}~\bibnamefont {Schneider}},
  \bibinfo {author} {\bibfnamefont {O.~J.}\ \bibnamefont {Boisseau}}, \bibinfo
  {author} {\bibfnamefont {P.}~\bibnamefont {Haase}}, \bibinfo {author}
  {\bibfnamefont {E.}~\bibnamefont {Slooten}}, \ and\ \bibinfo {author}
  {\bibfnamefont {S.~M.}\ \bibnamefont {Dawson}},\ }\href@noop {} {\bibfield
  {journal} {\bibinfo  {journal} {Behavioral Ecology and Sociobiology}\
  }\textbf {\bibinfo {volume} {54}},\ \bibinfo {pages} {396} (\bibinfo {year}
  {2003})}\BibitemShut {NoStop}%
\bibitem [{\citenamefont {Gleiser}\ and\ \citenamefont
  {Danon}(2003)}]{gleiser2003community}%
  \BibitemOpen
  \bibfield  {author} {\bibinfo {author} {\bibfnamefont {P.~M.}\ \bibnamefont
  {Gleiser}}\ and\ \bibinfo {author} {\bibfnamefont {L.}~\bibnamefont
  {Danon}},\ }\href@noop {} {\bibfield  {journal} {\bibinfo  {journal}
  {Advances in complex systems}\ }\textbf {\bibinfo {volume} {6}},\ \bibinfo
  {pages} {565} (\bibinfo {year} {2003})}\BibitemShut {NoStop}%
\bibitem [{\citenamefont {K{\"o}tter}(2004)}]{kotter2004online}%
  \BibitemOpen
  \bibfield  {author} {\bibinfo {author} {\bibfnamefont {R.}~\bibnamefont
  {K{\"o}tter}},\ }\href@noop {} {\bibfield  {journal} {\bibinfo  {journal}
  {Neuroinformatics}\ }\textbf {\bibinfo {volume} {2}},\ \bibinfo {pages} {127}
  (\bibinfo {year} {2004})}\BibitemShut {NoStop}%
\bibitem [{\citenamefont {Bock}\ \emph {et~al.}(2011)\citenamefont {Bock},
  \citenamefont {Lee}, \citenamefont {Kerlin}, \citenamefont {Andermann},
  \citenamefont {Hood}, \citenamefont {Wetzel}, \citenamefont {Yurgenson},
  \citenamefont {Soucy}, \citenamefont {Kim},\ and\ \citenamefont
  {Reid}}]{bock2011network}%
  \BibitemOpen
  \bibfield  {author} {\bibinfo {author} {\bibfnamefont {D.~D.}\ \bibnamefont
  {Bock}}, \bibinfo {author} {\bibfnamefont {W.-C.~A.}\ \bibnamefont {Lee}},
  \bibinfo {author} {\bibfnamefont {A.~M.}\ \bibnamefont {Kerlin}}, \bibinfo
  {author} {\bibfnamefont {M.~L.}\ \bibnamefont {Andermann}}, \bibinfo {author}
  {\bibfnamefont {G.}~\bibnamefont {Hood}}, \bibinfo {author} {\bibfnamefont
  {A.~W.}\ \bibnamefont {Wetzel}}, \bibinfo {author} {\bibfnamefont
  {S.}~\bibnamefont {Yurgenson}}, \bibinfo {author} {\bibfnamefont {E.~R.}\
  \bibnamefont {Soucy}}, \bibinfo {author} {\bibfnamefont {H.~S.}\ \bibnamefont
  {Kim}}, \ and\ \bibinfo {author} {\bibfnamefont {R.~C.}\ \bibnamefont
  {Reid}},\ }\href@noop {} {\bibfield  {journal} {\bibinfo  {journal} {Nature}\
  }\textbf {\bibinfo {volume} {471}},\ \bibinfo {pages} {177} (\bibinfo {year}
  {2011})}\BibitemShut {NoStop}%
\bibitem [{\citenamefont {Adamic}\ and\ \citenamefont
  {Glance}(2005)}]{adamic2005political}%
  \BibitemOpen
  \bibfield  {author} {\bibinfo {author} {\bibfnamefont {L.~A.}\ \bibnamefont
  {Adamic}}\ and\ \bibinfo {author} {\bibfnamefont {N.}~\bibnamefont
  {Glance}},\ }in\ \href@noop {} {\emph {\bibinfo {booktitle} {Proceedings of
  the 3rd international workshop on Link discovery}}}\ (\bibinfo {organization}
  {ACM},\ \bibinfo {year} {2005})\ pp.\ \bibinfo {pages} {36--43}\BibitemShut
  {NoStop}%
\bibitem [{\citenamefont {Guimera}\ \emph {et~al.}(2003)\citenamefont
  {Guimera}, \citenamefont {Danon}, \citenamefont {Diaz-Guilera}, \citenamefont
  {Giralt},\ and\ \citenamefont {Arenas}}]{guimera2003self}%
  \BibitemOpen
  \bibfield  {author} {\bibinfo {author} {\bibfnamefont {R.}~\bibnamefont
  {Guimera}}, \bibinfo {author} {\bibfnamefont {L.}~\bibnamefont {Danon}},
  \bibinfo {author} {\bibfnamefont {A.}~\bibnamefont {Diaz-Guilera}}, \bibinfo
  {author} {\bibfnamefont {F.}~\bibnamefont {Giralt}}, \ and\ \bibinfo {author}
  {\bibfnamefont {A.}~\bibnamefont {Arenas}},\ }\href@noop {} {\bibfield
  {journal} {\bibinfo  {journal} {Physical review E}\ }\textbf {\bibinfo
  {volume} {68}},\ \bibinfo {pages} {065103} (\bibinfo {year}
  {2003})}\BibitemShut {NoStop}%
\bibitem [{\citenamefont {Watts}\ and\ \citenamefont
  {Strogatz}(1998)}]{watts1998collective}%
  \BibitemOpen
  \bibfield  {author} {\bibinfo {author} {\bibfnamefont {D.~J.}\ \bibnamefont
  {Watts}}\ and\ \bibinfo {author} {\bibfnamefont {S.~H.}\ \bibnamefont
  {Strogatz}},\ }\href@noop {} {\bibfield  {journal} {\bibinfo  {journal}
  {nature}\ }\textbf {\bibinfo {volume} {393}},\ \bibinfo {pages} {440}
  (\bibinfo {year} {1998})}\BibitemShut {NoStop}%
\bibitem [{\citenamefont {Spring}\ \emph {et~al.}(2002)\citenamefont {Spring},
  \citenamefont {Mahajan},\ and\ \citenamefont
  {Wetherall}}]{spring2002measuring}%
  \BibitemOpen
  \bibfield  {author} {\bibinfo {author} {\bibfnamefont {N.}~\bibnamefont
  {Spring}}, \bibinfo {author} {\bibfnamefont {R.}~\bibnamefont {Mahajan}}, \
  and\ \bibinfo {author} {\bibfnamefont {D.}~\bibnamefont {Wetherall}},\ }in\
  \href@noop {} {\emph {\bibinfo {booktitle} {ACM SIGCOMM Computer
  Communication Review}}},\ Vol.~\bibinfo {volume} {32}\ (\bibinfo
  {organization} {ACM},\ \bibinfo {year} {2002})\ pp.\ \bibinfo {pages}
  {133--145}\BibitemShut {NoStop}%
\bibitem [{\citenamefont {Papadopoulos}\ \emph {et~al.}(2012)\citenamefont
  {Papadopoulos}, \citenamefont {Kitsak}, \citenamefont {Serrano},
  \citenamefont {Bogun{\'a}},\ and\ \citenamefont
  {Krioukov}}]{papadopoulos2012popularity}%
  \BibitemOpen
  \bibfield  {author} {\bibinfo {author} {\bibfnamefont {F.}~\bibnamefont
  {Papadopoulos}}, \bibinfo {author} {\bibfnamefont {M.}~\bibnamefont
  {Kitsak}}, \bibinfo {author} {\bibfnamefont {M.~{\'A}.}\ \bibnamefont
  {Serrano}}, \bibinfo {author} {\bibfnamefont {M.}~\bibnamefont {Bogun{\'a}}},
  \ and\ \bibinfo {author} {\bibfnamefont {D.}~\bibnamefont {Krioukov}},\
  }\href@noop {} {\bibfield  {journal} {\bibinfo  {journal} {Nature}\ }\textbf
  {\bibinfo {volume} {489}},\ \bibinfo {pages} {537} (\bibinfo {year}
  {2012})}\BibitemShut {NoStop}%
\end{thebibliography}%

\end{document}